\documentclass[fleqn,usenatbib]{mnras}

\usepackage[utf8]{inputenc}

\usepackage{graphicx}
\usepackage{lscape}
\usepackage{float}
\usepackage{epsfig}
\usepackage{multirow}

\usepackage{bigdelim}
\usepackage{bigstrut}

\def\aap{Astronomy \& Astrophysics}

\title[S-PLUS: LEnticular Galaxies in Stripe 82]{S-PLUS: LEnticular Galaxies in Stripe 82}
\author[Cortesi et al.]{A. Cortesi$^{1}$\thanks{E-mail: aricorte@astro.ufrj.br}, K. Saha$^{2}$, F.Ferrari$^{3}$, G. Lucatelli$^{3}$, C. Mendes de Oliveira$^{4}$, S. Dhiwar$^{2}$,
\newauthor
C. R. Bom$^{5}$, L. O. Dias$^{5}$\\
$^{1}$ Observat\'orio do Valongo, Ladeira (OV)\\
$^{2}$  Inter-University Centre for Astronomy and Astrophysics (IUCAA)\\
$^{3}$  Universidade Federal do Rio Grande (FURG)\\
$^{4}$ Universidade de S\~{a}o Paulo \\
$^{5}$ Brazilian Center for Physics Research (CBPF)}

\begin{document}

\date{Accepted in  Anais da Academia Brasileira de Ciencias
}

\pagerange{\pageref{firstpage}--\pageref{lastpage}} \pubyear{2021}

\maketitle

\label{firstpage}

\begin{abstract}
This work is a Brazilian-Indian collaboration. It aims at investigating the structural properties of Lenticular galaxies in the Stripe 82 using a combination of S-PLUS (Southern Photometric Local Universe Survey) and SDSS data. S-PLUS is a novel optical multi-wavelength survey which will cover nearly 8000 square degrees of the Southern hemisphere in the next years and the first data release covers the Stripe 82 area. The morphological classification and study of the galaxies' stellar population will be performed combining the Bayesian Spectral type (from BPZ) and Morfometryka (MFMTK) parameters. 
BPZ and MFMTK are two complementary techniques, since the first one determines the most likely stellar population of a galaxy, in order to obtain its photometric redshift (phot-z), and the second one recovers non-parametric morphological quantities, such as asymmetries and concentration. The combination of the two methods allows us to explore the correlation between galaxies shapes (smooth, with spiral arms, etc.) and their stellar contents (old or young population). The preliminary results, presented in this work,  show how this new data set opens a new window on our understanding of the nearby universe.
\end{abstract}


\section{Introduction}
Lenticular galaxies have been largely studied, since when they have been first classified by Edwin Hubble~\cite{Hubble:1936}, as the missing link between elliptical and spiral galaxies. S0s or lenticular galaxies present, in fact, an hybrid structure, showing a prominent disk, but with no signs of spiral arms. The presence of the disk suggests that their kinematics is rotationally supported~\cite{Cortesi:13a,Cortesi:13b}, as found for spiral galaxies~\cite{Merrett:03}. Elliptical galaxies, on the contrary, are pressure supported systems, as a result of their formation histories, which comprise a sequence of major and minor mergers~\cite{Bournaud:05}. Moreover, lenticular galaxies are more frequently found in denser regions of galaxy clusters~\cite{Dressler:80} and their frequency increases towards low redshifts~\cite{Dressler:97}. Considering that this is an opposite trend to that of spiral galaxies, it has been first hypothesised that S0 galaxies may be the end product  of spiral galaxies whose gas was gently stripped or consumed. Such evolutionary path would require the interplay between the galaxy and the surrounding environment to cease star formation and remove the spiral arms. Yet, field lenticular galaxies exist. At the present moment, several mechanisms have been pointed out as possible ways to create lenticular galaxies: minor mergers~\cite{Bournaud:05}, major mergers under specific initial conditions~\cite{Eliche-Moral:18}, AGN feedback or/and secular evolution~\cite{Mishra:18}, and it is becoming more popular the idea that the S0 class is actually a compilation of objects that formed through very different paths and just happen to have  similar appearances. Part of the problem stands on the difficulty of visually identifying lenticular galaxies. In fact, there is not a consensus on the morphological characteristics of lenticular galaxies and they can, in general, be easily misclassified due to the vague appearence of spiral arms, and not trivial determination of the presence of a disk, from a purely photometric approach. Large spectroscopic surveys, on the other side, often covers only the central part of a galaxy, i.e. the region where the bulge dominates, again making it hard to separate elliptical galaxies, from lenticular galaxies with large bulge-over-total light ratio. Multi-wavelength surveys (as S-PLUS~\cite{Mendes:19}) open a new window on galaxy classification, combining the possibility of studying the variation of morphological parameters with wavelengths and recovering the galaxy stellar population. We present preliminary results of this approach, using the S-PLUS first data release\footnote{https://datalab.noao.edu/splus/}, and we describe new methods to identify lenticular galaxies.

\section{Data and Methods}\label{previous work}
\begin{figure}
\includegraphics[width=\columnwidth]{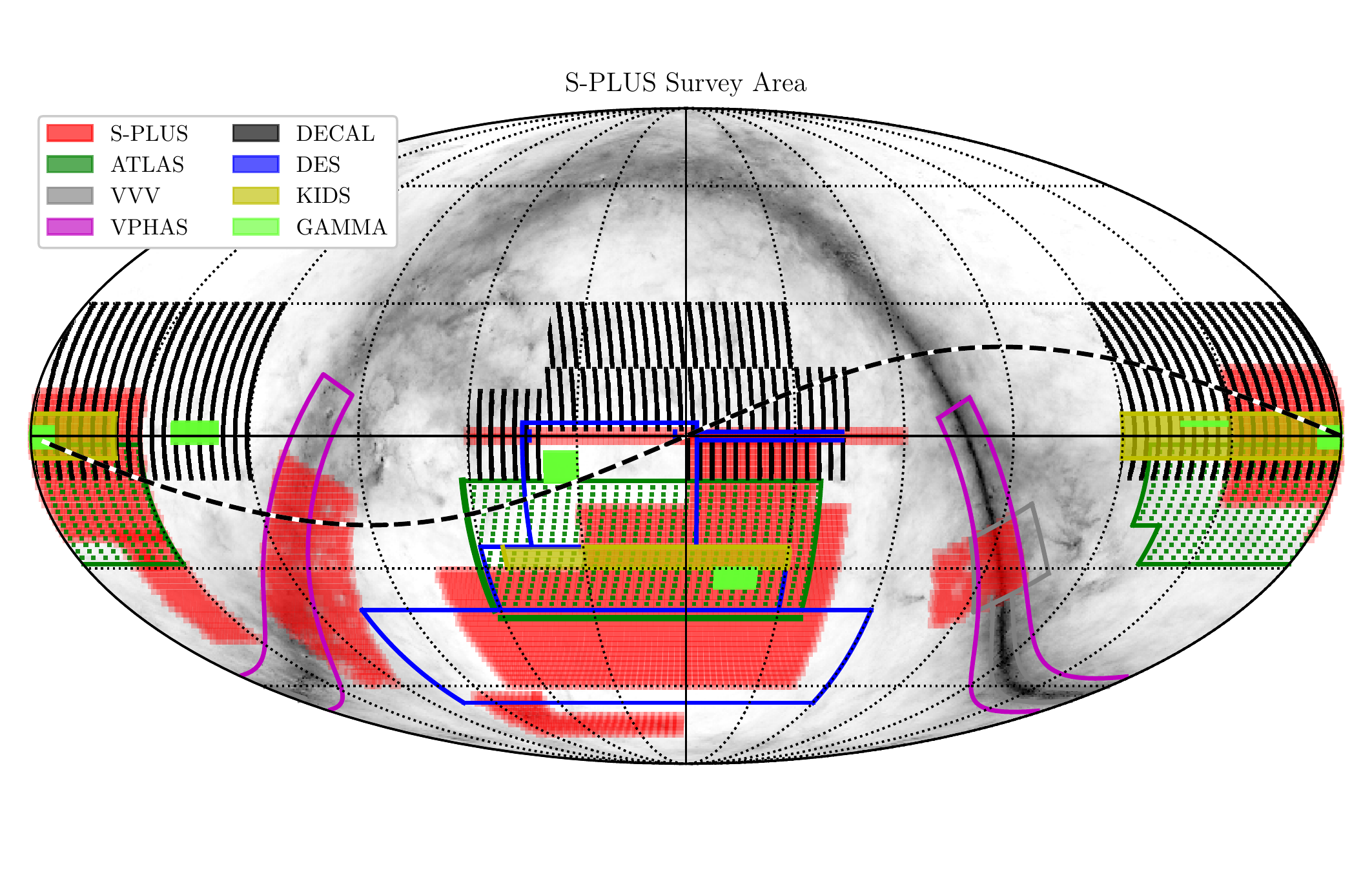}
\caption[ESO-SPLUS synergies]{Diagram showing some of the most important optical and near-infrared surveys in the Southern Hemisphere (we omit the surveys SkyMapper, Gaia and LSST that cover the entire hemisphere or sky). For the optical surveys: ATLAS is shown in hatched green, VPHAS+ is the pink rectangular contour over the Bulge and Disk of the Galaxy, DECAL is in hatched black, DES is shown in blue,  KiDS in yellow, and GAMA in bright green. The only near-infrared survey displayed is VISTA-VVV, a rectangle with light gray contours over the Galactic disk.   Dark red shows the area covered by S-PLUS. The broken black line represents the ecliptic. The figure is taken from~\cite{Mendes:19}.}
\label{fig:sky}
\end{figure}

\begin{figure}
\includegraphics[width=0.8\columnwidth,angle=-90]{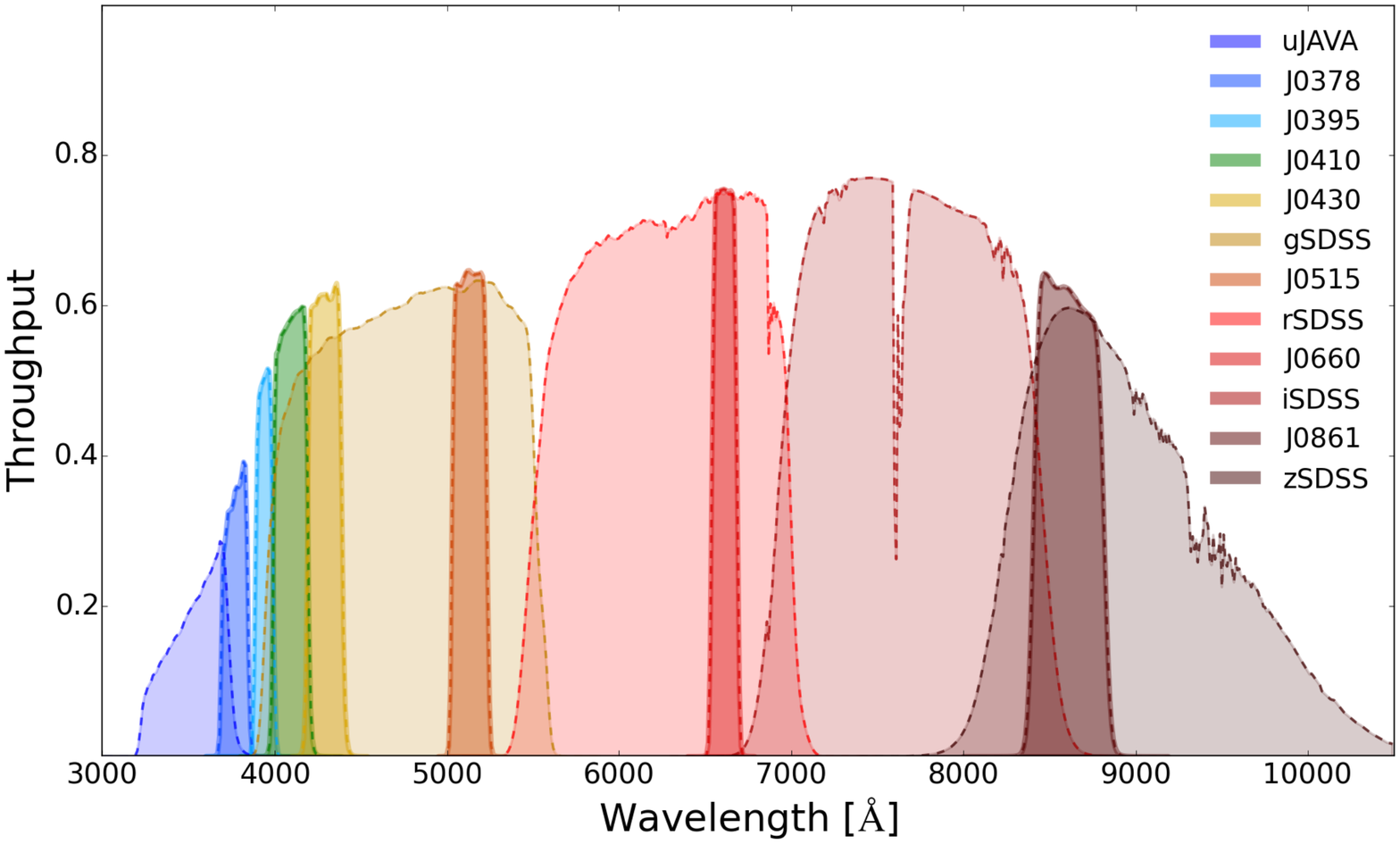}
\caption[The Javalambre 12-filter system used by S-PLUS]{The Javalambre 12-filter system used by S-PLUS. The y-axis shows the total efficiency of the filters, obtained through the multiplication of the average transmission curves, the CCD efficiency, and the mirror reflectivity curves. Different filters are coloured according to the labels on the right-side panel. The figure is taken from~\cite{Mendes:19}.}
\label{fig:filtersystem}
\end{figure}

S-PLUS (Southern Photometric Local Universe Survey) is an astronomical facility in Chile (Cerro Pachon), dedicated to mapping the observable sky in 7 narrow-band filters and 5 broad-band (ugriz) filters in the optical region. The 0.86m mirror of the T80-South telescope, combined with a field of view of 1.4 square degrees and an 85 Mega-pixel camera, is producing high-quality images and a unique spectral resolution for millions of objects over several thousand square degrees. Together with its twin observatory in the northern hemisphere, the T80-North, its sister survey J-PLUS, will cast the first light on a multi-colour Universe, on about  half of the extragalactic sky. This multi-purpose astrophysical survey in the southern hemisphere started in 2016. During the next 3-4 years it will observe more than 8,000 square degrees (1/5 of the whole sky), covering the entire visible region of the electromagnetic spectrum (3500 A to 10,000 A). Figure \ref{fig:sky} and \ref{fig:filtersystem} show the area covered in the sky and the Javalambre filter system used by S-PLUS.

Galaxy morphologies vary with wavelength, since different stellar populations present different colours. In general, star forming galaxies have a lower bulge-to-total ($B/T$) light ratio than quiescent ones~\cite{Morselli:17}.
We use \textsc{Morfometryka}~\cite{Ferrari:15} (MFMTK) to compute galaxies'  non-parametric morphometric parameters, in particular we recover the concentration (C1) and the entropy (H).  The concentration  is defined as  the ratio of the radii containing some fraction of the total light inside the Petrosian Region (2Rp): 
\[
C_1 = \log \left( \frac{ R_{80} }{ R_{20} } \right) \qquad,
\]
where $R_f$ is the radius that contains a fraction of $f$ \%.
Entropy is defined as:
\[
H(I)= -\sum_{i}^{n}p(I_{i})\ln[p(I_{i})]/H_{max},
\]
where $p$ is the probability of the occurence of the intensity $I_{i}$ and $H_{max}$ is the maximum entropy, which is for an homogeneous distribution (e.g. $p=1/n$). 
For the early type galaxies the concentration  is nearly constant with wavelength, while it increases for redder wavelengths, in the case of the spiral galaxies.
Another feature is that homogeneous systems (with less concentrated light distributions, such as disks or extended galaxies) have higher entropies than systems with more concentrated light distributions (ellipticals, bulge dominated, compacted, etc.).
Finally, we use the Bayesian Photometric redshifts (BPZ)~\cite{Molino:14} method to obtain the galaxy bayesian spectral type, by identifying the galaxy template model that optimises the redshift determination, for more details see~\cite{Molino:20}.

\section{Results}\label{results}

\begin{figure}
\includegraphics[width=\columnwidth]{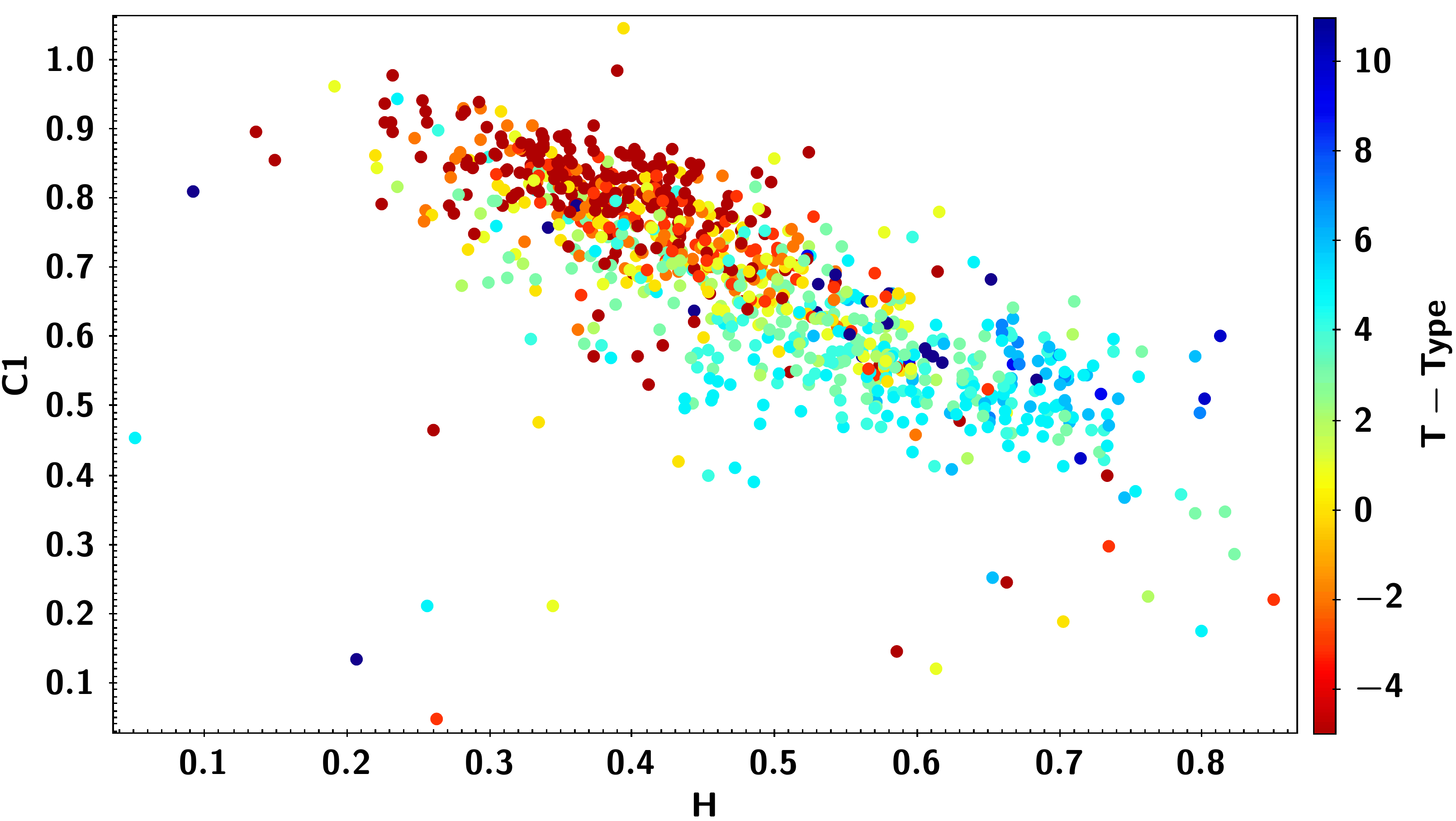}\\
\includegraphics[width=\columnwidth]{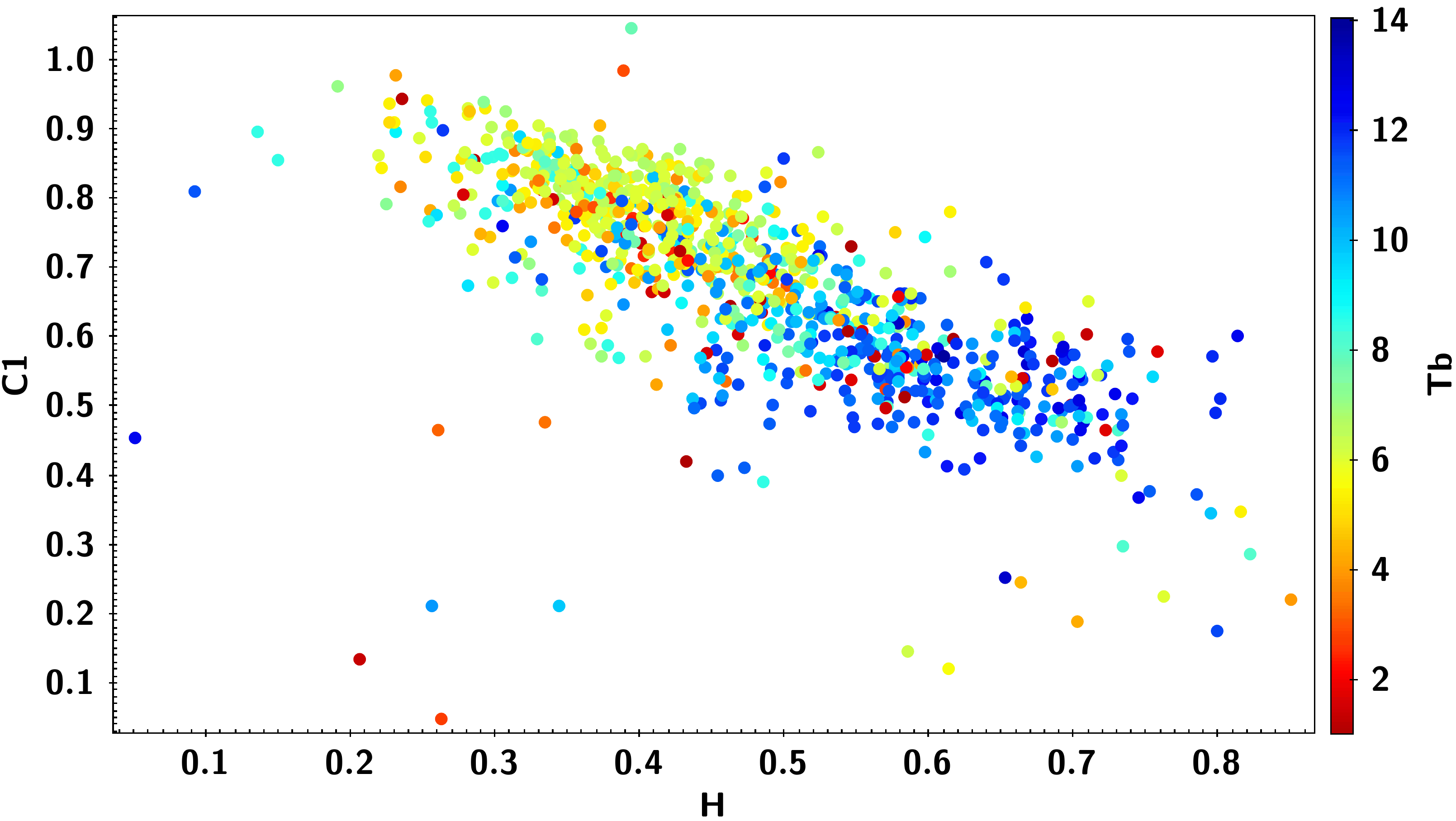}
\caption[H-C1 diagrams]{Entropy versus concentration parameters, estimated using MFMTK, for a sample of 925 galaxies, colour coded according to their morphological T-Type, top panel, and according  to their bayesian spectral templates, bottom panel. See text for more details.   }
\label{fig:result}
\end{figure}

The top panel of Figure \ref{fig:result} shows the distribution of the galaxies in the $H-C1$ plane, colour coded according to their morphological type~\cite{Nair:10}. The T-Type parameter value is lower than $-3$ for elliptical galaxies, between $-3$ to $0$ for lenticular galaxies, and higher than $0$ for spiral and irregular galaxies. This plot clearly shows that only using these two MFMTK~\cite{Ferrari:15} parameters we are able to separate galaxies into different classes. Combining these parameters with other MFMTK outputs (such as asymmetries, clumpiness, Gini parameter, spirality ..) and the curvature of the brightness profile of the galaxies~\cite{Lucatelli:19} at every wavelengths, we are able to identify a bona-fide catalogue of lenticular galaxies for the S-PLUS DR1 data release (Lucatelli et al. in prep). 

The bottom panel of Figure \ref{fig:result} presents the same $H-C1$ relation, but this time the objects are colour coded according to their Bayesian Spectral Type (Tb), as defined in~\cite{Molino:20}.  The galaxies' templates are defined as follows: from T1 to T4  is a spectral energy distribution (SED) typical of elliptical galaxies, T5 is for ES0 galaxies, from T6 to T14 are spiral galaxies templates (barred and un-barred). Again, the morphometric parameters allow us to divide galaxies into passive (early-type galaxies) and star forming (late-type galaxies). Interestingly, several T1 galaxies lie in the part of the plot populated by spiral galaxies, and some galaxies characterised by a spiral-like galaxy templates fall in the upper area of the plot, where we expect to find early type galaxies. This finding shows that galaxies' morphology and stellar content are not always as expected: the universe present blue ellipticals and red spirals~\cite{Strateva:01}. S-PLUS data set allows to recover both these quantities and map the galaxy distribution in the local universe.

\section{Conclusions}\label{conclusions}

In this work we present preliminary results of the study of galaxy morphology and stellar population properties obtained using data of the S-PLUS survey. We show how the parameters recovered with MFMTK allow to perform a galaxy morphological classification and we suggest a way to identify S0 galaxies, combining such parameters with a novel method based on the study of the curvature of the galaxies' surface light profile~\cite{Lucatelli:19}. The S-PLUS data set, given its unique combination of narrow and broad band filters, allows the estimation of the galaxy most probable stellar template, together with the definition of its morphometric parameters. This study extends from observational astronomy, to theory and computational analysis; it comprises researchers from Brazil and India and it is open to a world-wide collaboration. With its 0.86 cm diameter, S-PLUS proves the importance of small telescopes in the advance of astronomical knowledge. Several other small telescopes exist in the BRICS countries. We conclude encouraging a dialogue among these telescope users, sharing technical and scientific knowledge. 

\section{Acknowledgments}\label{acknowledgments}
AC acknowledges support from PNPD/CAPES.

\label{lastpage}
\bibliographystyle{mnras}
\bibliography{legs82}




\end{document}